\documentclass[twocolumn,showpacs,preprintnumbers,amsmath,amssymb]{revtex4}

\usepackage{graphicx}
\usepackage{dcolumn}
\usepackage{bm}
\newcommand{\YSO}{Y$_2$SiO$_5$}
\newcommand{\PrYSO}{$Pr^{3+}$:$Y_2SiO_5$}

\newcommand{\bbang}{\textit{Bang Bang}}
\newcommand{\CP}{\textit{Critical Point}}

\begin{document}
\preprint{APS/123-QED}

\title{Dynamic decoherence control of a solid-state nuclear quadrupole
  qubit}

\author{E. Fraval}
\email{elliot.fraval@anu.edu.au}
\author{M. J.  Sellars}
\author{J. J. Longdell}

\affiliation{Laser Physics Center, Research School of Physical
  Sciences and Engineering, Australian National University.}

\date{\today}

\begin{abstract}

  We report on the application of a dynamic decoherence control pulse
  sequence on a nuclear quadrupole transition in \PrYSO\ . Process
  tomography is used to analyse the effect of the pulse sequence. The
  pulse sequence was found to increase the decoherence time of the
  transition to over 30 seconds. Although the decoherence time was
  significantly increased, the population terms were found to rapidly
  decay on the application of the pulse sequence. The increase of
  this decay rate is attributed to inhomogeneity in the ensemble.
  Methods to circumvent this limit are discussed.

\end{abstract}

\maketitle

Recent experiments have demonstrated that optically active centers in
solids and in particular rare-earth ions in crystals show promise as
an alternative to atomic based ensembles for quantum information
processing applications\cite{rotpaper,PhaseGate}.  Rare-earth doped
ions provide multi-level systems possessing optical transitions with
long coherence times and in crystalline hosts, high optical densities
are achievable.  These solid state centers have the significant
advantage over atomic systems that they are stationary, thus allowing
spatial and frequency variation in an ensemble induced by interactions
with light fields to be maintained for appreciable times. This last
attribute has already been exploited extensively in the field of time
domain optical processing, where complex temporal and spatial
relationships between different interacting light fields are employed
to process classical information, in a manner not possible in atomic
systems\cite{tian01}.

A common feature of many of the proposed ensemble based quantum
information processing applications, including quantum memories,
single photon sources and quantum repeaters, involves the storage of
quantum information on long lived low lying
states\cite{juls04,kuhn02,brie98}.  In the case of rare-earth centers
it is proposed to store the quantum information on the centers' ground
state hyperfine transitions\cite{ohls02,mois01}. The duration and
fidelity that such information can be stored depends on the coherence
time of the hyperfine transitions and, therefore, it is important to
maximize these coherence times.  The coherence times of hyperfine
transitions in rare-earth centers are typically of the order of ms but
in recent work we have demonstrated ways where these can be extended
and have achieved a coherence time of $82ms$ in the case of
$m_I=-1/2\leftrightarrow+3/2$ transition in \PrYSO\ \cite{CP}. In this
letter we demonstrate a further dramatic increase of the coherence
time to several tens of seconds through the use of a \bbang\ dynamic
decoupling sequence.

Dynamic decoupling methods for open systems were initially developed
for use in NMR spectroscopy to selectively remove contributions to the
spin Hamiltonian\cite{call91,hahn50}. There is growing interest in
applying these techniques and in particular \emph{``Bang Bang''}
decoupling sequences to quantum information systems to decrease their
rate of decoherence \cite{viol98,viol99,viol04,byrd02,vital99}.
\emph{``Bang Bang''} sequences operate by decoupling the quantum
system from the bath through the application of a periodic control
Hamiltonian. In the current work this is implemented by a series of
pairs of hard pulses $\left(\pi,-\pi\right)$ separated by the cycling
time $\tau_c$. The \bbang\ pulse sequence can theoretically rephase
all the coherence in the quantum system, thereby making $T_2=T_1$ if
the following criterion\cite{viol98} is met.
\begin{equation}
  \label{eq:BangBangCondition}
  \omega_c\tau_c \lesssim 1
\end{equation}
where $\omega_c$ is the cutoff frequency of the dephasing bath and
$c\tau_c$ is the period between the $\pi$ pulses in the \bbang\ 
sequence. The pulses are assumed to be `hard' such that during the
pulse any evolution of the state other than the action of the driving
field can be assumed to be negligible.

For Pr ions substituting for Y in \YSO\ the main source of decoherence
of the its ground state hyperfine transitions is magnetic interactions
with nuclei in the host\cite{CP,Pr-Y}. Y possess a nuclear spin and
has a magnetic moment of $\gamma_Y$$=$$209Hz/G$. The other nuclear
spin in the host is $^{29}Si$ found in natural abundance ($\sim$4\%)
with a moment of $845Hz/G$. The magnetic field seen by any given Pr
ion fluctuates over time due to resonant cross relaxations between the
host spins. The sensitivity can be calculated using the reduced
Hamiltonian of the electronic ground state hyperfine
structure\cite{rotpaper}. Praseodymium has a nuclear spin of 5/2 and
as a result of a pseudoquadrupole interaction its electronic ground
state splits into three doubly levels with splittings of the order of
10 MHz. The final degeneracy can be lifted by a magnetic field through
an enhance nuclear Zeeman interaction. In the region where the applied
field produces anticrossings a field direction and magnitude can be
found such that there is no first order Zeeman shift for the $m_I=-1/2
\leftrightarrow +3/2$ hyperfine transition, demonstrated in recent
work\cite{CP}. It has been shown that the dephasing perturbations
acting on the Pr ions are predominantly occurring on a timescale
longer than $\sim$$10ms$\cite{CP}. Given that Rabi frequencies greater
than $100kHz$ are readily achievable the period between the applied
pulses in the \bbang\ sequence can be made such that the inequality (1)
is satisfied for the frequencies of the dominant perturbations.

\begin{figure}
  \includegraphics[width=0.41\textwidth]{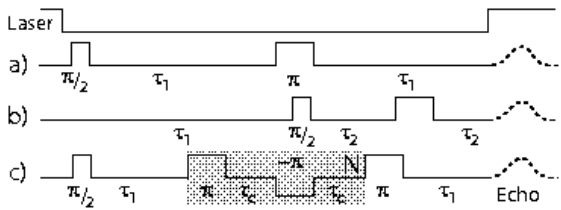}
\caption{\label{fig:PulseSequence} Pulse sequences used in
experiment: a) 2 pulse spin echo, b) Inversion Recovery $T_1$ c)
\bbang\ pulse sequence. The coherence generated by the initial
pulse is rephased by a pair of $\pi,-\pi$ pulses separated by the
cycling time $\tau_c$, iterated N times. The laser is common to
all pulse sequences}
\end{figure}
\begin{figure}
  \includegraphics[width=0.25\textwidth]{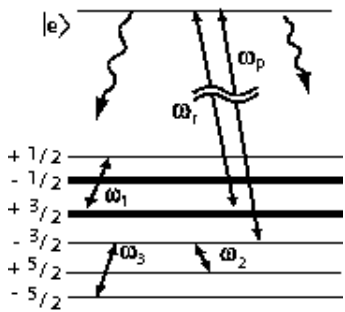}
\caption{\label{fig:repump} All ground state hyperfine levels
other than $m_I=-1/2$ interact via RF frequencies $\omega_{1-3}$
with laser radiation either at the read frequency $\omega_r$ or
pump frequency $\omega_p$. Through spontaneous emission from the
electronic excited state the group of ions interacting with both
$\omega_r$ and $\omega_p$ via any common excited state will be
holeburned into the $m_I=-1/2$ state. Bold states are the \CP\
transition.}
\end{figure}

As in previous work a \PrYSO\ $0.05\%$ concentration crystal
maintained at temperature of $\sim$$1.5K$ in a liquid helium bath
cryostat\cite{CP}. Yttrium orthosilicate (\YSO\ ) is a low symmetry
host with two crystallographically inequivalent sites where Pr can
substitute for Y, labelled `site 1' and `site 2'\cite{equa95}. Site 1
ions are used in this work.  The coordinate system chosen is the $C_2$
axis is $y$, $z$ is the direction of the predominate polarization of
the optical ${}^3\!H_{4}-{}^1\!D_2$ transition and $x$ is
perpendicular to both.

The magnetic fields to achieve the \CP\ field configuration
($\bm{B_{\text{CP}}}=\left(732,173,-219\right)G$\cite{CP}) were
supplied by two orthogonal superconducting magnets supplying a $z$
field, and $x,y$ field. The sample was rotated about the $z$ axis
to provide the correct ratio of fields along the $x$ and $y$ axes
for the critical point in magnetic field space. The field in the
$x,y$ plane could also be adjusted using a small correction coil
mounted orthogonal to the main $x,y$. The inhomogeneity in
magnetic field across the sample was measured using a hall probe
to be $<$$2G$.

Raman heterodyne was employed to investigate the ground state
hyperfine transitions using an experimental setup similar to that
described in previous work\cite{CP}. The experiment was performed
using a Coherent 699 frequency stabilized tunable dye laser tuned to
the ${}^3\!H_{4}-{}^1\!D_2$ transition at 605.977nm. The laser's
frequency was stabilized to a sub kilohertz linewidth. The laser power
incident on the crystal was $40mW$, focused to $\sim$$100 \mu m$ and
could be gated using a $100MHz$ acousto-optic modulator.  The
hyperfine transition was excited using a six turn coil with a diameter
of $5mm$, driven by a $10W$ RF amplifier resulting in a Rabi frequency
$\Omega_rf=91kHz$. The RF pulse and digital control sequences were
generated using a direct digital synthesis system referenced to an
oven controlled crystal oscillator. The pulse sequences used in the
experiment are illustrated in Fig. \ref{fig:PulseSequence} The Raman
heterodyne signal, seen as a beat on the optical beam, was detected by
a $125MHz$ photodiode.  This signal was analyzed using a mixer and a
phase controlled local oscillator referenced to the RF driving field.

At the \CP\ field the $m_I=-1/2 \leftrightarrow +3/2$ transition
was observed at 8.646 MHz with a inhomogeneous linewidth of
$4kHz$. This linewidth was found to be insensitive to changes in
the magnetic field of the order of $\sim$$10G$.

Prior to applying each Raman heterodyne pulse sequence the sample was
prepared with the optical/RF repump scheme as shown in Fig.
\ref{fig:repump}. The repump frequencies were
$\omega_r-\omega_p=18.2MHz$, $\omega_{1}=12.2$, $\omega_{2}=15.35MHz$
and $\omega_{3}=16.3MHz$. The repump RF was pulsed with a duty cycle
of 10\% to reduced sample heating, while the repump laser frequency
$\omega_r$ was scanned 200kHz to hole burn a trench in the
inhomogeneous optical line where detection would take place. This
repump scheme ensures that all Pr ions interacting with the laser
radiation are forced into the $m_I=-1/2$ state, creating a pure state
ensemble. It also ensures there is no initial population near the
laser frequency used for Raman heterodyne detection. The use of a sub
kilohertz linewidth laser and the repump scheme resulted in a
significant improvement in the signal to noise compared to work
performed earlier\cite{CP}.

\begin{figure}
  \includegraphics[width=0.45\textwidth]{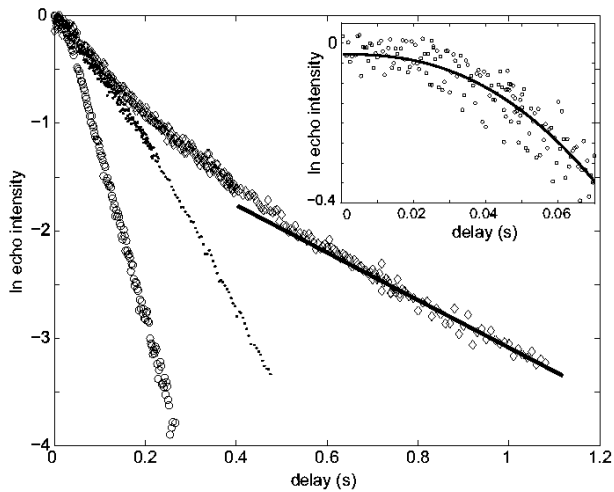}
  \caption{\label{fig:2Pulse}Spin echo decay at the \CP\ field ($\diamond$), and detuned from the \CP\ field by $\sim$2G ($\cdot$) and 5G ($\circ$) in the Z direction)}
\end{figure}

Figure \ref{fig:2Pulse} shows the decay of the amplitude of two
pulse spin echoes as a function of the delay between the pulses.
Two data sets are shown, the first is for an applied magnetic
field optimized for long decay times and the second for a field
detuned by $5G$ in the $z$ direction from the optimal value. The
significantly longer decay time for the optimised field in the
present work, compared to previous work\cite{CP} is attributed to
better identification of the critical point enabled by the
addition of the correction coil enabling significantly more
precise magnetic field adjustments. Besides being longer the decay
can no longer be described by standard echo decay function with a
single time constant\cite{mims68}. There are three distinct
regions.  For pulse separations less than $20ms$ the decay rate is
less than $1/4 s^{-1}$.  At $30ms$ there is a distinct shoulder
with the decay rate increasing to $1/0.4 s^{-1}$ as the pulse
seperation reaches $60ms$. From 150 ms onwards the decay rate
asymptotically approaches a value of $1/0.86 s^{-1}$. This
asymptotic decrease in the decay rate was only observed for
magnetic fields within $0.5G$ of the optimal field.  When the
field was more than $0.5G$ away from the critical point a simple
exponential decay was observed for delays longer than 50 ms.

The shoulder in the decay at $30ms$ is interpreted as indicating
that the majority of the dephasing is due to perturbations that
occur on time scales between 10 and $100ms$. The asymptotic
behavior of the decay is attributed to a variation in the $T_2$
within the ensemble resulting from inhomogeneity in the magnetic
field across the sample. Ions experiencing a field closer to the
\CP\ condition will have a longer $T_2$ and consequently their
contribution to the echo intensity will dominate for long pulse
separations. The asymptotic decay rate therefore gives an upper
limit for the contribution to the decoherence rate due to second
order magnetic interactions.

\begin{figure}
  \includegraphics[width=0.41\textwidth]{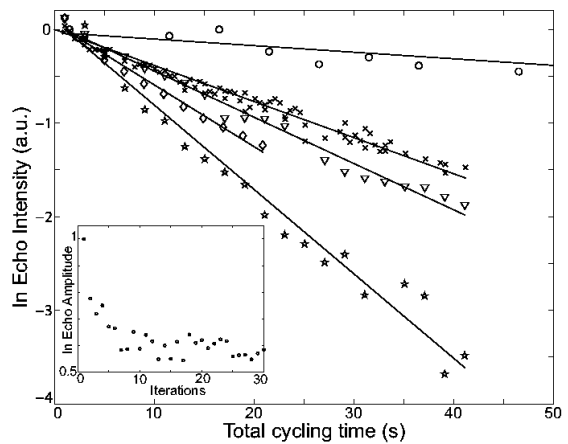}
\caption{\label{fig:bangbang} \bbang\ decoupled echo decays with
  $\tau_c=7.5ms(\star), 10ms(\diamond), 15ms(\nabla), 20ms(\times)$
  corresponding to $T_2= 27.9, 21.1, 15.2, 10.9s$. Inversion recovery
  measurements ($\circ$) yield $T_1=145s$}
\end{figure}

\begin{figure}
  \includegraphics[width=0.41\textwidth]{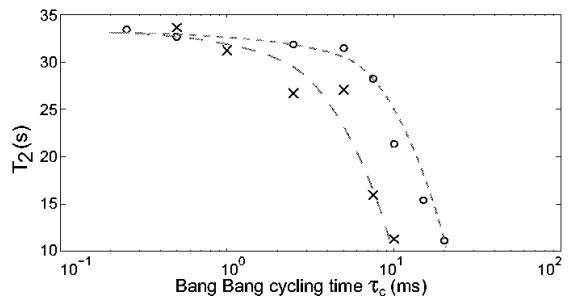}
\caption{\label{fig:bangbangComparison} Dependence of decoherence
  time on the \bbang\ cycling time $\tau_c$ both at the critical point
  ($\circ$) and with the magnetic field misaligned to give a coherence
  time of $T_2 = 100ms$ ($\times$). Trend lines are included but do
  not represent a physical model.}
\end{figure}

The \bbang\ pulse sequence was investigated using an initial delay of
$\tau_1=1.2ms$ and varying the cycling time $\tau_c$ from $20ms$ to
$0.5ms$, as shown in Fig. \ref{fig:bangbang}. Also shown in Fig.
\ref{fig:bangbang}is the result of an inversion recovery measurement
used to determine the lifetime of the transition. The inversion
recovery measurements were performed using the pulse sequence
described in Fig. \ref{fig:PulseSequence}(b). As can be seen in Fig.
\ref{fig:bangbang} the \bbang\ sequence significantly increases $T_2$,
though for the shortest value of $\tau_c$ $T_2/T_1<1/4$.  In Figure
\ref{fig:bangbangComparison} the coherence times $T_2$ for each of the
data sets are plotted as a function of the cycling time $\tau_c$ .
This shows that for $\tau_c < 5 ms$, there are significant gains in
$T_2$ made by reducing $\tau_c$, while reducing $\tau_c$ further only
slightly increases in $T_2$. Also shown in Figure
\ref{fig:bangbangComparison} is the same measurements made with the dc
magnetic field detuned so as to increase the transitions magnetic
field sensitivity. The field was detuned such that the two pulse echo
$T_2$ was reduced to 100ms.  Although for long $\tau_c$ there is a
significant reduction in the $T_2$ observed using the \bbang\ for the
measurements made with the detuned field, $T_2$ in the limit of short
$\tau_c$ appears to be the same for the two magnetic field conditions.
This suggest that the residual decoherence in the limit of short
$\tau_c$ is not predominately due to magnetic field fluctuations.  The
residual decoherence is possibly due to detuning and pulse area errors
in the decouplng sequence.
\begin{figure}
  \includegraphics[width=0.41\textwidth]{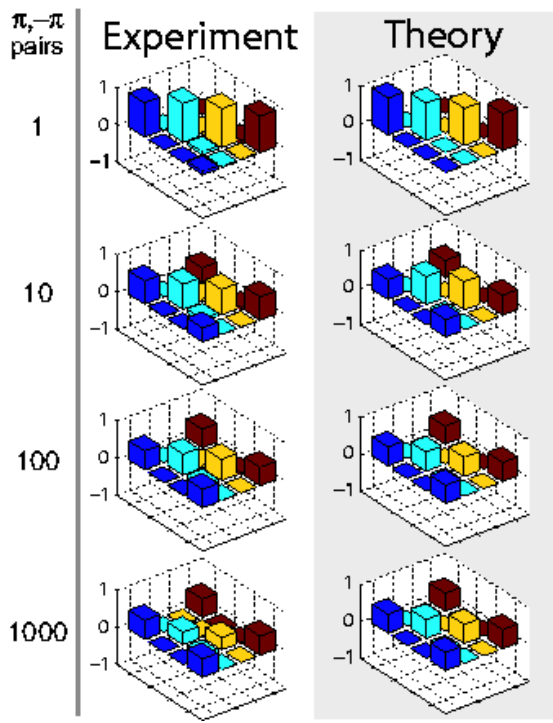}
\caption{\label{fig:tomog} Process tomography of the \bbang\ pulse
  sequence for 1, 10, 100 and 1000 iterations for both experiment and
  theoretical modeling. The imaginary component of both the experiment
  and modeling is always zero and is omitted for clarity}
\end{figure}

To assess how well arbitrary quantum state are preserved by the
\bbang\ pulse sequence we performed process tomography on the input
state and for 1, 10, 100 and 1000 iterations of the \bbang\ cycle. The
initial delay was $\tau_1=1.2ms$, with a cycling time of $tau_c=2ms$.
The total period over which the tomography was performed was
$\sim$$4ms$, $\sim$$40ms$, $\sim$$400ms$ and $\sim$$4s$ respectively.
The imaginary component of the process tomography is only shown for
the input state since it never contributed more than 10\% of the
signal.  Ideally the \bbang\ process operator is the identity matrix,
leaving the state unchanged. It was observed that the component of the
Bloch vector in the coherence plane for a given state was preserved
well, while the population component of the Bloch vector rapidly
decayed. The fidelity of the \bbang\ process for 1, 10, 100 and 1000
iterations was 99\%, 65\%, 54\% and 43\% respectively.

The evolution of the ensemble was modeled using the Bloch equations
assuming an infinite $T_1$ and $T_2$, with an inhomogeneous linewidth
of $4 kHz$ (FWHM) and a Rabi frequency of $100 kHz$. The results from
this modeling are shown in figure \ref{fig:tomog}, along side the
experimental data. Despite the model not including any homogeneous
dephasing the main features of the simulated process tomography
(figure \ref{fig:tomog}) closely matches those of the experimental
data. In particular the rapid decay of the population component of the
Bloch vector compared to the coherence components.  Further
simulations indicated that the decay rate of the population terms can
be reduced by increasing the ratio of the Rabi frequency to the
linewidth. A suitable criteria for when the application of the \bbang\ 
pulse sequence is useful for preserving arbitrary quantum states is
when the decay rate of the population terms in the presence of the
pulse sequence is slower than that of the coherence terms in the
absence of the sequence.  For the present case where $T_2=0.86s$ the
simulation indicates that to meet this criteria it will be necessary
to achieve a ratio of Rabi frequency to linewidth of
$\Omega_{RF}/\omega_{inh}$$\approx$$100$. There is limited capacity to
increase the Rabi frequency of the driving field without the possible
excitation of off-resonant transitions.  Therefore for the application
of the \bbang\ pulse sequence to the $m_I=-1/2 \leftrightarrow +3/2$
transition to be useful it will be necessary to reduce the
inhomogeneous broadening of the transition by a factor of $\sim$10.

The large change in magnetic field sensitivity as the \CP\ field is
approached with no corresponding change in the inhomogeneous linewidth
of the transition suggests that the inhomogeneous broadening of the
transition at the \CP\ field is not due to magnetic interactions.  The
inhomogeneous broadening at the \CP\ field is probably due to strain
within the crystal which is not intrinsic to the site and can be
reduced by refining standard crystal growing techniques. A reduction
in strain broadening by over an order of magnitude has been achieved
in analagous materials\cite{macf98} through reducing the dopant
concentration. Irrespective of reaching the desired ratio of Rabi
frequency to inhomogeneous linewidth, methods of designing pulse
sequences more robust to Rabi and detuning errors have been
proposed\cite{viol03,wocj04}.

In conclusion the application of the \bbang\ pulse sequence
demonstrates that dynamic decoupling techniques are applicable to
correcting quantum errors on nuclear spin transitions of \PrYSO .
This work realises very long hyperfine decoherence times, greater than
30s, in a solid state optical $\Lambda$ system suitable for quantum
information processing applications.

The support of this work by the Defence Science Technology
Organization (DSTO) and the Australian Research Council (ARC) is
gratefully acknowledged.

\bibliography{/home/elf111/Documents/papers/writing/complete.bib}

\end{document}